\documentclass[%
 reprint,
 amsmath,amssymb,
 aps,
]{revtex4-1}

\usepackage{graphicx}
\usepackage{dcolumn}
\usepackage{bm}

\begin{document}

\preprint{APS/123-QED}

\title{Deriving the properties of space time using the non-compressible solutions of the Navier-Stokes Equations}
\author{Ryan McDuffee}
\email{rmcduffe@du.edu}

\date{\today}
\begin{abstract}
Recent observations of gravitational waves by the Laser Interferometer Gravitational-Wave Observatory (LIGO) has confirmed one of the last outstanding predictions in general relativity and in the process opened up a new frontier in astronomy and astrophysics. Additionally the observation of gravitational waves has also given us the data needed to deduce the physical properties of space time. Bredberg et al have shown in their 2011 paper titled From Navier-Stokes to Einstein, that for every solution of the incompressible Navier-Stokes equation in p + 1 dimensions, there is a uniquely associated “dual" solution of the vacuum Einstein equations in p + 2 dimensions. The author shows that the physical properties of space time can be deduced using the recent measurements from the Laser Interferometer Gravitational-Wave Observatory and solutions from the incompressible Navier-Stokes equation.
\end{abstract}

\pacs{Valid PACS appear here}
\maketitle


\section{\label{sec:level1}Introduction}

Albert Einstein first published his theory of general relativity in 1915, a year later he would predict the existence of gravitational waves \cite{abbott2016observation}. He based this prediction on his observation that the linearized weak field equations had wave solutions \cite{einstein1916naherungsweise}\cite{einstein1918gravitationswellen}, these transverse waves would be comprised of spatial strain and travel at the speed of light and would be generated by the time variations of the mass quadrupole moment of the source. Einstein also observed that the amplitude of these waves would be incredible small and thus he never expected to observe them \cite{abbott2016observation}. That same year Karl Schwarzschild would publish his solutions to the Einstein field equations that would predict the existence of black holes and in 1963 he would generalize these solutions to rotating black holes\cite{schwarzschild1916gravitational}\cite{kerr1963gravitational}. 
\par In 2016 the Laser Interferometer Gravitational-Wave Observatory(LIGO) announced they had detected a gravitational wave on September 14, 2015 at 09:50:45 UTC, The signals frequency ranged from 35 to 250 Hz with a peak gravitational-wave strain of \begin{math}1*10^{-21}\end{math}\cite{abbott2016observation}. This wave matched the waveform predicted by general relativity for the in spiral and merger of a pair of black holes and the ringdown of the resulting single black hole. 
\par The detection of such a wave is important on its own because the measurement of the waveform and amplitude of the gravitational waves from a black hole merger (such as the one observed by LIGO) event makes possible accurate determination of its distance. The accumulation of black hole merger data from cosmologically distant events may help to create more precise models of the history of the expansion of the universe and the nature of the dark energy that influences it. However the direct detection of gravitational waves also allows use to determine the physical properties of space, by using solutions common to both the incompressible Navier-Stokes equation and the Einstein field equations of general relativity to calculate a hypothetical “viscosity” of space time and thus deduce the structure.
\par Before moving forward with the analysis, it must be understood that the “Viscosity” of space time that the author is intending to calculate, is an analogues property rather than a literal one. Meaning that rather then trying to estimate the “thickness” of a fluid like space time, we intend to show that by calculating the viscosity term in a Navier-Stokes like equation, it is possible to calculate a Shear modulus that describes the deformation of space time, thus allowing for the analysis of space time at its most basic level. Before this can be done though, it is necessary to understand how the solutions of the Navier-Stokes equation relate to the solutions of the Einstein field equation and how the data from the 2016 observations at the Laser Interferometer Gravitational-Wave Observatory (LIGO) can be used to calculate derive the hypothetical properties of space time.

\section{\label{sec:level2}The Navier-Stokes equation and Einstein field equation}
Both the Einstein field equation \begin{math}G_{\mu\nu}=0\end{math} and the Navier-Stokes equation \begin{math}\dot{v_{i}}-\eta\partial^2v_{i}+\partial_iP+v^{j}\partial_jv_i=0 \end{math} have long been incredibly powerful tools for understanding the world both terms of mathematics and physics. The Einstein equation universally governs the long-distance behavior of gravitational systems, while the incompressible Navier-Stokes equation universally governs the hydrodynamic limit of essentially any fluid. Both equations have an intricate non-linear structure that allow them to be applied to wide verity of systems and thus offer us an insight into a wide range of physical phenomena. Before these two equation can be used to probe the structure of space time though it necessary to understand that for that for every solution of the incompressible Navier-Stokes equation in p+ 1 dimensions, there is a uniquely associated “dual" solution of the vacuum Einstein equations in p + 2 dimensions\cite{bredberg}. Those interested in a more rigorous demonstration of this assertion should read From Navier-Stokes to Einstein by Dr. Irene Bredberg et al\cite{bredberg}. 
\par What follows is very brief overview their work, the structure of both the Einstein field equation and Navier-Stokes, and the boundary conditions necessary for there to be a solution to each equation. We begin by seeking a relation between the (p+2)-dimensional Einstein and (p+1)-dimensional Navier-Stokes equations. Since, the former has a much larger solution space than the latter, only a special type of Einstein geometry is relevant. Roughly speaking, the relevant geometries are non-singular perturbations of a horizon. A more precises description of the geometries we are interested in can be found and From Navier-Stokes to Einstein by Dr. Irene Bredberg et al \cite{bredberg} for the moment we are only interested in the boundary conditions of these geometries which we will denote as \begin{math}\sum_{c}\end{math}. The boundary hypersurface \begin{math}\sum_{c}\end{math} is taken to be asymptotically null in both the far future and far past. In Minkowskian coordinates \begin{math}ds^2_{p+2}=-dudv+dx_idx^i\end{math}, past null infinity \begin{math}\mathcal{I}^-\end{math} is the union of the null surface \begin{math}v \rightarrow -\infty\end{math} together with \begin{math}u \rightarrow -\infty\end{math} and \begin{math}\sum_{c}\end{math} is a time like hyper surface \begin{math}uv=-4r_{c}\end{math} with \begin{math}v>0\end{math}\cite{bredberg}. Past (future) event horizons \begin{math}\mathcal{H}^- \mathcal{H}^+\end{math} are defined by the boundaries of the causal future (past) of \begin{math}\sum_{c}\end{math} near-horizon and the hydrodynamic \begin{math}\epsilon\end{math}-expansion\cite{bredberg}. 
\par These two expansions are shown to be equivalent \begin{math}\eta=r_c\end{math}. Initial data can be specified on the union of \begin{math}\sum_{c}\end{math} and \begin{math}\mathcal{I}^-\end{math}. We consider initial data which is asymptotically Minkowskian and flat (no incoming waves) on \begin{math}\mathcal{I}^-\end{math}. On \begin{math}\sum_{c}\end{math} we generally demand that the intrinsic metric \begin{math}\gamma_{ab}\end{math} be flat meaning that \begin{math}\gamma_{ab} = \eta_{ab}\end{math} for \begin{math}ab=0,...p\end{math}. Next we wish to consider the general solution of the Einstein equations consistent with this initial data and smooth on \begin{math}\mathcal{H}^+\end{math} In particular. 
\par So far we have not specified the extrinsic curvature \begin{math}K_{ab}\end{math} on \begin{math}\sum_{c}\end{math}  or equivalently (and more conveniently) the Brown-York stress tensor on \begin{math}\sum_{c}\end{math} this will give us the equation \begin{equation}
  T \equiv 2 \gamma_{ab}K-K_{ab}  
\end{equation}\cite{bredberg}. If no initial data were prescribed on \begin{math}\mathcal{I}^-\end{math}, any \begin{math}T_{ab}\end{math} on \begin{math}\sum_{c}\end{math} consistent with the constraint equations could be chosen. This data could then in general be evolved radially inwards to produce a space time everywhere inside of \begin{math}\sum_{c}\end{math}\cite{bredberg}.In general, such a space time will have gravitational flux (if not singularities) going up to \begin{math}v = \infty (\mathcal{I}^+)\end{math} as well as down to \begin{math}\mathcal{I}^-\end{math}.Hence we have a “shooting problem" to find those special allowed choices of \begin{math}T_{ab}\end{math} which produce a space time smooth on \begin{math}\mathcal{H}^+\end{math} with no flux coming from \begin{math}\mathcal{I}^+\end{math}\cite{bredberg}. 
\par Dr. Bredberg has devised a complete solution to this problem in \cite{bredberg2011wilsonian}. We then implement Dr. Bredberg solution to leading order in a double expansion in long wavelengths and weak fields. Ingoing Rindler coordinates were used for which the leading order at metric is 
\begin{equation}
  ds^2_{p+2}=-rd\tau^2+2d\tau dr+dx_idx^i  
\end{equation} 
here \begin{math}\sum_{c}\end{math} is the accelerated surface\begin{math}r=r_c\end{math}, \begin{math}\mathcal{H}^-\end{math} is \begin{math}\tau=-\infty\end{math}, and \begin{math}\mathcal{H}^+\end{math} is r=0\cite{bredberg}. These coordinates are convenient for analyzing smoothness on \begin{math}\mathcal{H}^+\end{math}. It was found that the allowed choices of \begin{math}T_{ab}\end{math} are precisely those corresponding to the linearized fluid: \begin{math}r^{3/2}_cT^{\tau i}=v_i, r^{3/2}_cT^{ij}= -\eta \partial^{ij}\end{math}\cite{bredberg}. Where the (kinematic) viscosity here is given by the formula \begin{math}\eta = r_c\end{math} while \begin{math}v_i\end{math} obeys the linearized incompressible Navier-Stokes equation \begin{math}\partial_iv^i=0, \partial_{\tau}v^i-\eta\partial^2v_i=0\end{math}\cite{bredberg}. If we choose any value for the viscosity other than \begin{math}\eta = r_c,\end{math} the constraint equations on \begin{math}\sum_{c}\end{math} are still obeyed, but gravitational waves are propagated down to \begin{math}\mathcal{I}^-\end{math} and there is a singularity at \begin{math}r=0\end{math}\cite{bredberg}. Dr. Bredberg solution goes one step further and solves the problem in certain hydrodynamic and near-horizon limits without making a linearized approximation, enabling us to see a direct connection between the nonlinear structures of the Navier-Stokes and Einstein equations\cite{bredberg}.
\par We can now use a modified version of the Wilsonian Approach to Fluid/Gravity in order to solve the shooting problem in the long wavelength perturbation of \begin{math}\sum_{c}\end{math} without a simultaneous linearized expansion. The general solution will be parametrized by \begin{math}\epsilon\end{math} of the full nonlinear Navier-Stokes equation with viscosity \begin{math}\eta = r_c\end{math}\cite{bredberg}. First we must consider the metric 
\begin{equation}
 ds^2=-rd\tau^2+2d\tau dr+dx_idx^i    
\end{equation}
\begin{equation*}
    -[2(1-\frac{r}{r_{c}})v_{i}dx^{i}d\tau+(1-\frac{r}{r_{c}})(\partial_{j}v^{i}+\partial_{i}v^{j})dx^{i}dx^{j}
\end{equation*}
\begin{equation*}
    -2(r-\frac{r^{2}}{2r_{c}}-\frac{r_{c}}{2})\partial^{2}v_{i}dx^{i}d\tau ]
\end{equation*}
\begin{equation*}
 -\delta(\tau-\tau_{*})[(4(1-\frac{r}{r_{c}})F_{i}+\frac{2\alpha_{i}}{r_{c}})dx^{i}d\tau-\frac{2\beta_{ij}}{r_{c}}dx^{i}dx^{j}+...] 
\end{equation*}
\cite{bredberg}, where \begin{math}F_i\end{math} is an arbitrary function of \begin{math}x_i\end{math} obeying \begin{math}\partial_iF^i=0\end{math}. \begin{math}\beta_{ij}\end{math} and \begin{math}\alpha_{i}\end{math} (which is divergence free) are both functions of \begin{math}x^i\end{math} and related to \begin{math}F_i\end{math} by 
\begin{equation}
  \partial^i\partial_j\alpha_i=F_i  
\end{equation}, 
\begin{equation}
    \beta_{ij}=\partial_i\alpha_j+\partial_j\alpha_i
\end{equation}\cite{bredberg}. Since the metric on \begin{math}\sum_{c}\end{math} is no longer flat, the constraint equations become linearized Navier-Stokes equations with \begin{equation}
\partial_\tau v^i-\eta\partial^2v^i=F^i(x)\delta(\tau-\tau_*)
\end{equation}.
Since \begin{math}v_i(x,\tau )\end{math} is taken to vanish for \begin{math}\tau>\tau_*\end{math} the forcing term will cause it to jump to \begin{math}F_i(x^i)\end{math} at \begin{math}\tau=\tau_*\end{math} after which it will evolve according to Navier-Stokes. Given \begin{equation}
v^i-\eta\partial^2v^i=F^i(x)\delta(\tau-\tau_*)
\end{equation}\cite{bredberg} this geometry solves the linearized Einstein equations everywhere, and is characterized by an arbitrary divergence-free vector field \begin{math}F_i(x)\end{math}. Before at \begin{math}\tau=\tau_*\end{math} it is flat while afterward it is, up to a coordinate transformation, the linearization of 
\begin{equation}
    ds^2_{p+2}=-rd\tau^2+2d\tau dr+dx_idx^i
\end{equation}. At the nonlinear level, the equations are cumbersome and we have been unable to explicitly construct the analog of equation 3
away from \begin{math}
\sum_{c}\end{math}\cite{bredberg}. However it seems plausible that qualitatively similar solutions persist at the nonlinear level.
\section{\label{sec:level3}The 2016 observation of Gravitational Waves}
On September 14th 2015, the Laser Interferometer Gravitational-Wave Observatory (LIGO) directly observed gravitational waves from the inward spiral and merger of a pair of black holes of around 36 and 29 solar masses \cite{abbott2016observation}. Up until then existence of gravitational waves had only been inferred indirectly, through their effect on the timing of pulsars in binary star systems \cite{weisberg2004relativistic}. The observation of these waves gives us the information we need to use the Navier-Stokes equation to calculate the hypothetical viscosity of space time, primarily it allows us to calculate the frequency and amplitude of the gravitational wave. This calculation was possible prior to the detection of Gravitational waves, however with the direct observation of the waves, we can now use data that is physically meaningful.
\par The two black holes involved in the merger had masses of approximately 36(+5, -4) and 29(\begin{math}\pm\end{math} 4) solar masses\cite{abbott2016observation}. The recorded power of the gravitational wave peaked at \begin{math}3.6*10^49\end{math} watts and Across the 0.2-second duration of the detectable signal, the relative tangential (orbiting) velocity of the black holes increased from 3 percent to 6 percent of the speed of light\cite{abbott2016observation}. The orbital frequency of 75 Hz (half the gravitational wave frequency) means that the objects were orbiting each other at a distance of only 350 km by the time they merged\cite{abbott2016observation}. With this information we can now calculate the amplitude of the gravitational wave. To do this though, we must first derive the wave polarization equations for the system.
\par We begin with the equation for the space time curvature expressed with respect to a covariant derivative \begin{math}\bigtriangledown\end{math}, in the form of the Einstein tensor \begin{math}G_{\mu \nu }\end{math}. We can then relate the curvature to the stress–energy tensor \begin{math}T_{\mu \nu }\end{math}, with the equation \begin{math}G_{\mu \nu }=\frac{8\pi G_n}{c^4}T_{\mu \nu }\end{math}\cite{schutz2009first}. Where \begin{math}G_n\end{math} is Newton's gravitational constant, and c is the speed of light, for the purpose of this paper we will use geometrized units so that \begin{math}G_n= 1 = c\end{math} in order to simplify our calculations. We can now rewrite the Einstein's equations as wave equations and define our flat space, our flat space time will be given as 
\begin{equation*}
\eta_{\mu \nu }= \begin{bmatrix}-1&0&0&0\\0&1&0&0\\0&0&r^2&0\\0&0&0&r^2 \sin^2\theta  \end{bmatrix}
\end{equation*} (it should be noted that in the next section \begin{math}\eta\end{math}\cite{schutz2009first} will also be used to calculate our hypothetical “viscosity” of space time.) This flat-space metric has no physical significance; it is a purely mathematical device necessary for the analysis. 
\par Now, we can also think of the physical metric \begin{math}g_{\mu \nu }\end{math} as a matrix and find its determinant det g. Finally we define our radiation field in terms of our flat space time \begin{math}g_{\mu \nu }\end{math} and find the determinant det g, this will give us 
\begin{equation}\overline{h}^{\alpha\beta}  \equiv \eta ^{\alpha \beta }-g^{\alpha \beta } \sqrt{ \begin{vmatrix}\det(g)\end{vmatrix}}
\end{equation}. Next we define out coordinates in such a way that this quantity satisfies the de Donder gauge condition (conditions on the coordinates): \begin{math}\bigtriangledown_\beta  \overline{h}^{\alpha \beta }=0\end{math}. Where \begin{math}\bigtriangledown\end{math} represents the flat-space derivative operator. These equations say that the divergence of the field is zero. The linear Einstein equations can now be written as \begin{math}\Box  \overline{h}^{\alpha \beta }=16\pi \tau ^{\alpha \beta }\end{math}\cite{RevModPhys.52.299} where  \begin{math}\Box =-\partial^2_t + \bigtriangleup\end{math} is the flat-space d'Alembertian operator, and \begin{math}\tau ^{\alpha \beta }\end{math} represents the stress–energy tensor plus quadratic terms involving \begin{math}\overline{h}^{\alpha \beta }\end{math}\cite{RevModPhys.52.299}.This is just a wave equation for the field with a source, despite the fact that the source involves terms quadratic in the field itself. That is, it can be shown that solutions to this equation are waves traveling with velocity 1 in these coordinates. 
\par In order to obtain a numerical result from this equation though it will need to be linearized. We assume that space is nearly flat, so the metric is nearly equal to the \begin{math}\eta ^{\alpha \beta }\end{math} tensor. This means that we can neglect terms quadratic in \begin{math}\overline{h}^{\alpha\beta}\end{math}, which means that the \begin{math}\tau ^{\alpha \beta}\end{math} field reduces to the usual stress–energy tensor \begin{math}T ^{\alpha \beta}\end{math} and Einstein's equations become \begin{math}\Box  \overline{h}^{\alpha \beta }=16\pi T ^{\alpha \beta }\end{math}.However since we are interested in the field far from a source, we can treat the source as a point and everywhere else, the stress–energy tensor would be zero, so our equation becomes \begin{math}\Box  \overline{h}^{\alpha \beta }=0\end{math}. Now, this is just the usual homogeneous wave equation for each component of \begin{math}\overline{h}^{\alpha \beta}\end{math}. For a wave moving away from a point source, the radiated part (meaning the part that dies off as 1/r far from the source) can always be written in the form \begin{math}\frac{f(t-r,\theta ,\phi )}{r}\end{math}, where f is just a generic function. It can be shown that it is always possible to make the field traceless\cite{tredcr1975cw}. 
\par Now, if we further assume that the source is positioned at r=0, the general solution to the wave equation in spherical coordinates is \begin{equation*}\overline{h}^{\alpha \beta }= \frac{1}{r} \begin{bmatrix}0&0&0&0\\0&0&0&0\\0&0&h_{+}(t-r,\theta ,\phi )&h_{ \times }(t-r,\theta ,\phi )\\0&0&h_{ \times }(t-r,\theta ,\phi )&h_{+}(t-r,\theta ,\phi )\end{bmatrix}\end{equation*}\cite{tredcr1975cw}. Thus we have our polarization equations which can now be rewritten in Newtonian terms as 
\begin{equation}
h_{+}=\frac{-1}{R}\frac{G^2}{c^4}\frac{4m_1m_2}{r}
\end{equation}\cite{schutz2009first} and to calculate the amplitude of the gravitational wave detected by the Laser Interferometer Gravitational-Wave Observatory we simply substitute the measured values and get an upper value of \begin{math}25956/R\end{math} or a lower value of \begin{math}19890/R\end{math}, where R is the distance from the center of the system. The collision was detected approximately 1.3 billion Light years away from earth \cite{abbott2016observation}, and thus the spatial distortion caused by the gravitational wave was between \begin{math}2.1*10^{-21}\end{math} meters and \begin{math}1.6*10^{-21}\end{math} meters. This is the amplitude of the gravitational wave detected by the Laser Interferometer Gravitational-Wave Observatory and with the amplitude of the wave, we can now calculate the hypothetical properties of space time by using the Navier-Stokes equation.
\section{\label{sec:level4}The theoretical properties of space time as fluid}
We have now established that for that for every solution of the incompressible Navier-Stokes equation in p+ 1 dimensions, there is a uniquely associated “dual" solution of the vacuum Einstein equations in p + 2 dimensions \cite{bredberg} and calculated the amplitude of the wave 2016 gravitational wave. We also know that the gravitational wave detected by The Laser Interferometer Gravitational-Wave Observatory has velocity of c\cite{abbott2016observation}. With this information we can now use the Navier-Stokes equation to derive an equation for the velocity of wave in an incompressible medium and calculate the Shear modulus. We begin my considering a wave as disturbance of a fluid-like medium\cite{morse1953methods}\cite{chew1995waves}\cite{chung1978finite}. 
\par Next we write out the equations for the conservation of mass and momentum as \begin{equation}\frac{\partial \rho u_i}{\partial x_i}+\frac{\partial \rho}{\partial t}=0
\end{equation}
and 
\begin{equation}\frac{\partial \rho u_i}{\partial x_j} u_j+\frac{\partial \rho u_i}{\partial t}-\frac{\partial\sigma_{ij}}{\partial x_j}=0\end{equation}\cite{morse1953methods}\cite{chew1995waves}\cite{chung1978finite} where \begin{math}\rho\end{math} is the “mass density” (for our calculations this will be the smallest possible deformation in space time) u is the “particle” velocity, P is our fluid “pressure”. It should again be noted that when we refer to things like mass, particle and pressures, we are speaking metaphorically about analogous variables in continuum mechanics. The physical meaning of each of the analogous variables will be covered in the next section. We also have our three spatial coordinates are \begin{math}x_i (i=1,2,3)\end{math} for the domain \begin{math}\omega\end{math}. Particle velocities u are for each direction \begin{math}x_i\end{math}. Using the Navier-Stokes equation we can find the equation for incompressible fluid flow \begin{math}\rho \frac{du}{dt}=f_b- \bigtriangledown P\end{math} and the body forces are negligible giving us \begin{math}f_b=0\end{math}\cite{morse1953methods}.
\par The force \begin{math}f(t)\end{math} acts as a source function upon the domain\begin{math}\omega\end{math}. Any force acting within the domain causes pressure and density changes, and the fluid nature of the medium will create an equilibrium restoring force\cite{landau1954mechanics}.  Next we consider small perturbations \begin{math}\bigtriangleup\end{math} in the density \begin{math}\rho\end{math}, the particle velocity u, and the pressure P from the initial rest conditions which are labeled with subscript 0. \begin{math}u_t=u_0+ \bigtriangleup u\end{math}, \begin{math}\rho _t=\rho _0+ \bigtriangleup \rho\end{math},  \begin{math}P_t=P_0+ \bigtriangleup P\end{math}\cite{landau1954mechanics}. The initial particle velocity of our location in space is assumed to be 0 because the domain \begin{math}\omega\end{math} is assumed to be at rest relative to the particles around it. The density perturbation is based on the acoustic approximation\cite{landau1954mechanics}, in this approximation A fluid has a pressure which is a function of density, temperature, and gravitational forces. 
\par We shall assume that the gravity forces are relatively constant over the domain \begin{math}\omega\end{math} and do not exert any differential force on the fluid. We will neglect the effects of temperature as the changes are very small. Hence, we will assume that only the density is important and that the stress within the fluid is related to the strain as a function of density. Next we simply apply the Kronecker delta to our stress matrix \begin{equation*}\delta_{ij}= \begin{pmatrix}1&0&0\\0&1&0\\0&0&1\end{pmatrix}\end{equation*} and use Euler’s equation to get \begin{math}\rho \frac{du}{dt}=- \bigtriangledown P_t\end{math} and since the pressure is assumed to be constant the equation becomes \begin{math}
\rho \frac{du}{dt}=-\bigtriangledown \bigtriangleup P\end{math}\cite{artley1965fields}. Now the initial medium is at rest and has no convective acceleration, which permits changing the form of the derivatives and this gives us \begin{math}\rho \frac{\partial u}{\partial t}=- \bigtriangledown  \bigtriangleup P\end{math}.  Next u is the gradient of \begin{math}\phi\end{math}, and we can also see that the product of \begin{math}\bigtriangleup \rho\end{math} and the gradient of \begin{math}\phi\end{math} will be small thus \begin{math}\bigtriangledown \phi =u\end{math} and \begin{math}\rho_0\frac{\partial \bigtriangledown \phi }{\partial t}=- \bigtriangledown  \bigtriangleup P\end{math} \cite{morse1953methods}.
\par Next we assume the derivatives of time and space can be exchanged which will give us \begin{math}\rho_0\frac{\partial \phi }{\partial t}=- \bigtriangledown  \bigtriangleup P\end{math} and this equation can be simplified to \begin{math}\rho_0\frac{\partial \phi }{\partial t}=- \bigtriangleup P\end{math}. Next since we are measuring space time in terms of its deformation caused by waves, we use the shear modulus instead of the Young's modulus that would normally be used for calculating the velocity of a wave. To calculate the wave velocity, we take the incompressible fluid equation and apply mass conservation to the first derivative of the fluid equation and then solve for velocity which gives us \begin{math}u =  \sqrt{\frac{G}{\rho_0}}\end{math} (where G is the shear modulus). Next we calculate the shear modulus for space time \begin{math}G = \frac{Fl}{A \bigtriangleup x}\end{math} where F is the force of the two observed black holes colliding, l is the initial length of the smallest area effected by the force, A is the area over which the force acts (this would be the circular distance between earth and the colliding black holes) and \begin{math}\bigtriangleup x\end{math} would be the spatial displacement caused by the gravitational wave (the amplitude of the gravitational wave we calculated earlier). 
\par We now know all of the quantities for our velocity equation except for the initial length of the smallest area affected by the wave. By substituting the known quantities into the wave velocity equation and solving for l, we find that the smallest area affected by the observed gravitational wave was \begin{math}1.4*10^{-21}\end{math} meters.  We can now take this information and substitute our results into the equation for dynamic viscosity \begin{math}F = \mu A \frac{u}{y}\end{math} and solve for \begin{math}\mu\end{math} to get our hypothetical viscosity of space time of \begin{math}6.2616*10^{-39}\end{math} kg/ms. This for all intents and purposes gives us a substance with a viscosity that is functionally 0, and thus space time behaves as a perfect fluid.
\section{\label{sec:level5}Implications}
Before continuing to the consequences of space time behavior as a perfect/super fluid, let us first summarize our argument. We have already established that for every solution of the incompressible Navier-Stokes equation in p + 1 dimensions, there is a uniquely associated “dual" solution of the vacuum Einstein equations in p + 2 dimensions. This means that any solution derived from Navier-Stokes equation in p + 1 dimensions there is a “duel” solution of the vacuum Einstein equations in p + 2 dimensions. This establishes that our work in one set of equations is valid in the other assuming that both treat their respective mediums as incompressible and that a constant number of dimensions is used in each. 
\par We then take the polarization equation for a system of orbiting bodies and calculate the amplitude of the gravitational wave caused by this system using the values measured by the Laser Interferometer Gravitational-Wave Observatory in 2015 and then using a shear module and the equation for the velocity of a transvers wave in a medium derived from the Navier-Stokes equation and then solve for the initial length of the space prior to its distortion. We then take the force of the event that caused the spatial distortion, the initial size of the undistorted space, the area affected by the gravitational wave and then solve for the “viscosity” of space. The resulting “viscosity” tells us about how malleable space time is and as a result imposes constraints on the size of any extra dimensions space. In this context "viscosity" simply means the deformability of space time, our particles are simply quantized units of space, and the "pressure" is a uniform force present through out all of space time.
\par According to general relativity, the conventional gravitational wave is the small fluctuation of curved space time which has been separated from its source and propagates independently. In Superfluid vacuum theory, a subset of Grand Unified Field Theories where the fundamental physical vacuum (non-removable background) space time is viewed as a Superfluid and the curved space time we see in general relativity is the small collective excitation of a superfluid background\cite{zloshchastiev2009spontaneous}. Meaning that much like water, space time is composed of smaller parts. However the properties of these parts will vary widely depending on what formulation of Superfluid vacuum theory is being used and if Superfluid vacuum theory is being incorporated into another theory such as m-theory. However when treating space time as a super fluid and assuming that general relativity's interpretation of a gravitational wave is accurate, results in the following conclusions about the graviton. Primarily, the Graviton would be the "small fluctuation of the small fluctuation" meaning it would be the fundamental unit of space. This may not appear to be a physically robust concept as it is akin to trying to define a phonon in terms of smaller fluctuations inside it. However it may be possible to better ascertain the nature of such a strange phenomenon if we make the following assumptions about space times structure as a fluid.
\par First we assume that there are more than 3+1 dimension of space and that any extra dimensions of space can exist both at the quantum level and on the macroscopic level. Next we assume there are differences in the malleability of these dimensions of space time, meaning that higher dimensions are less malleable then the dimensions we experience every day. With these assumption in place, in theory we should be able to calculate the size of these extra dimension using the Einstein–Smoluchowski relation\cite{islam2004einstein}. However this is where we encounter our first problem with trying to deduce the size of these extra dimensions via the characteristics of gravitational wave. If we simply take the general case of the Einstein–Smoluchowski relation and calculate the mobility \begin{math}\mu\end{math} of our particles of space for the velocity of the expanding universe at a given distance (calculated using Hubble's law) over the force that caused the gravitational wave in conjunction with the ambient temperature of the vacuum, we can then solve for the diffusion constant, the above values give space time a diffusion constant of \begin{math}8.3097*10^{-57} m^2/s\end{math}. 
\par We then take the Stokes–Einstein equation\cite{mason2000estimating} and set it equal to the diffusion constant that we just calculated and then substitute our “viscosity” for n and then simply solve for r. If we do this then we get a radius of \begin{math}3.799*10^{70}m\end{math} to put this in perspective consider that the radius of the observable universe is \begin{math}4.4*10^{26} m\end{math}. This would seem to indicate that any extra dimension of space would have to be large extra dimensions akin to those in the Randall–Sundrum model\cite{davoudiasl2000bulk}. However we initially intended to calculate the size of compactified dimensions such as those in m-theory. However in order to do this we need to calculate the drift velocity of the individual units that in theory would comprise space time, however at this level, determining the exact the velocity of any individual unit becomes impossible due to the laws of quantum physics. Thus while our calculations on the macroscopic level introduce the intriguing possibility of large extra dimensions of space, in order to use the data measured by the Laser Interferometer Gravitational-Wave Observatory to deduce the nature of compactified extra dimensions of space, a more complete description of space time at the quantum level is required.
\bibliographystyle{unsrt}
\bibliography{Derivingthepropertiesofspacetime.bib}        

\begin{thebibliography}{10}

\bibitem{abbott2016observation}
BP~Abbott, Richard Abbott, TD~Abbott, MR~Abernathy, Fausto Acernese, Kendall
  Ackley, Carl Adams, Thomas Adams, Paolo Addesso, RX~Adhikari, et~al.
\newblock Observation of gravitational waves from a binary black hole merger.
\newblock {\em Physical review letters}, 116(6):061102, 2016.

\bibitem{einstein1916naherungsweise}
Wolfgang Trageser.
\newblock N{\"a}herungsweise integration der feldgleichungen der gravitation.
\newblock {\em Albert Einstein: Akademie-Vorträge}, pages 149--158, 2008.

\bibitem{einstein1918gravitationswellen}
Albert Einstein.
\newblock {\"U}ber gravitationswellen.
\newblock {\em Sitzungsberichte der K{\"o}niglich Preu{\ss}ischen Akademie der
  Wissenschaften (Berlin), Seite 154-167.}, 1918.

\bibitem{schwarzschild1916gravitational}
Karl Schwarzschild.
\newblock On the gravitational field of a point mass according to einstein’s
  theory.
\newblock {\em Sitzungsber. Preuss. Akad. Wiss., Phys. Math}, 1916.

\bibitem{kerr1963gravitational}
Roy~P Kerr.
\newblock Gravitational field of a spinning mass as an example of algebraically
  special metrics.
\newblock {\em Physical review letters}, 11(5):237, 1963.

\bibitem{bredberg}
Irene Bredberg, Cynthia Keeler, Vyacheslav Lysov, and Andrew Strominger.
\newblock From navier-stokes to einstein.
\newblock {\em Journal of High Energy Physics}, 2012(7):1--18, 2012.

\bibitem{bredberg2011wilsonian}
Irene Bredberg, Cynthia Keeler, Vyacheslav Lysov, and Andrew Strominger.
\newblock Wilsonian approach to fluid/gravity duality.
\newblock {\em Journal of High Energy Physics}, 2011(3):1--29, 2011.

\bibitem{weisberg2004relativistic}
Joel~M Weisberg and Joseph~H Taylor.
\newblock Relativistic binary pulsar b1913+ 16: thirty years of observations
  and analysis.
\newblock {\em arXiv preprint astro-ph/0407149}, 2004.

\bibitem{schutz2009first}
Bernard Schutz.
\newblock {\em A first course in general relativity}.
\newblock Cambridge university press, 2009.

\bibitem{RevModPhys.52.299}
Kip~S. Thorne.
\newblock Multipole expansions of gravitational radiation.
\newblock {\em Rev. Mod. Phys.}, 52:299--339, Apr 1980.

\bibitem{tredcr1975cw}
If~Tredcr et~al.
\newblock Cw misner, ks thorne, ja wheeler: Gravitation. wh freeman and company
  limited, reading (england) 1973—xxvi+ 1279 seiten, preis{\pounds} 19.20
  (clothbound);{\pounds} 8.60 (paperbound).
\newblock {\em Astronomische Nachrichten}, 296(1):45--46, 1975.

\bibitem{morse1953methods}
Philip~McCord Morse and Hermann Feshbach.
\newblock Methods of theoretical physics.
\newblock {\em Journal of Fluid Mechanics}, 1953.

\bibitem{chew1995waves}
Weng~Cho Chew.
\newblock {\em Waves and fields in inhomogeneous media}, volume 522.
\newblock IEEE press New York, 1995.

\bibitem{chung1978finite}
TJ~Chung.
\newblock Finite element analysis in fluid dynamics.
\newblock {\em NASA STI/Recon Technical Report A}, 78:44102, 1978.

\bibitem{landau1954mechanics}
LD~Landau and E~Mo Lifshits.
\newblock Mechanics of continuous media.
\newblock {\em Gostekhizdat, Moscow}, 81, 1954.

\bibitem{artley1965fields}
John Artley.
\newblock {\em Fields and configurations}.
\newblock Holt, Rinehart and Winston, 1965.

\bibitem{zloshchastiev2009spontaneous}
Konstantin~G Zloshchastiev.
\newblock Spontaneous symmetry breaking and mass generation as built-in
  phenomena in logarithmic nonlinear quantum theory.
\newblock {\em arXiv preprint arXiv:0912.4139}, 2009.

\bibitem{islam2004einstein}
MA~Islam.
\newblock Einstein--smoluchowski diffusion equation: a discussion.
\newblock {\em Physica Scripta}, 70(2-3):120, 2004.

\bibitem{mason2000estimating}
Thomas~G Mason.
\newblock Estimating the viscoelastic moduli of complex fluids using the
  generalized stokes--einstein equation.
\newblock {\em Rheologica Acta}, 39(4):371--378, 2000.

\bibitem{davoudiasl2000bulk}
H~Davoudiasl, JL~Hewett, and TG~Rizzo.
\newblock Bulk gauge fields in the randall--sundrum model.
\newblock {\em Physics Letters B}, 473(1):43--49, 2000.

\end{thebibliography}
\end{document}